\documentclass[12pt]{article}

\usepackage{graphicx}
\graphicspath{{./figs/}} 

\usepackage[letterpaper,margin=1in]{geometry}

\renewenvironment{abstract}
	{\quotation}
	{\endquotation}

\date{}

\makeatletter
\renewcommand{\fnum@figure}{\textbf{Figure \thefigure}}
\renewcommand{\fnum@table}{\textbf{Table \thetable}}
\makeatother

\usepackage{url}

\usepackage{subcaption}

\usepackage{amsmath,dsfont,amsfonts,braket}

\usepackage{hyphenat} 

\usepackage{microtype} 

\usepackage[dvipsnames]{xcolor}         
\usepackage[unicode=false,psdextra]{hyperref}
\hypersetup{
    pdfdisplaydoctitle,
    bookmarksnumbered=true,
    bookmarksopen,
    breaklinks,
    linktoc=all,
    plainpages=false,
    unicode=true,
    colorlinks=true,                  
    allcolors=black,     
    linkcolor=NavyBlue,      
    citecolor=NavyBlue
}

\usepackage[backend=biber,
    autocite=superscript,
    url=false,
    sorting=none,
    hyperref=true,
    isbn=false,
    citestyle=numeric-comp]{biblatex} 
\usepackage[noabbrev,nameinlink,capitalise]{cleveref} 

\crefformat{equation}{(#2#1#3)}
\crefrangeformat{equation}{(#3#1#4) to~(#5#2#6)}
\crefmultiformat{equation}{(#2#1#3)}%
{ and~(#2#1#3)}{, (#2#1#3)}{ and~(#2#1#3)}

\crefname{SI}{Supplementary Section}{Supplementary Section} 

\crefname{SubFig_a}{Figure}{Figures}
\crefname{SubFig_b}{Figure}{Figures}
\crefname{SubFig_c}{Figure}{Figures}
\crefname{SubFig_d}{Figure}{Figures}
\crefname{SubFig_e}{Figure}{Figures}
\crefname{SubFig_f}{Figure}{Figures}
\crefname{SubFig_bc}{Figures}{Figures}
\crefname{SubFig_cd}{Figures}{Figures}
\crefname{SubFig_ce}{Figures}{Figures}
\crefname{SubFig_ab}{Figures}{Figures}
\crefname{SubFig_ac}{Figures}{Figures}
\crefname{SubFig_df}{Figures}{Figures}

\creflabelformat{SubFig_a}{#2#1(a)#3}
\creflabelformat{SubFig_b}{#2#1(b)#3}
\creflabelformat{SubFig_c}{#2#1(c)#3}
\creflabelformat{SubFig_d}{#2#1(d)#3}
\creflabelformat{SubFig_e}{#2#1(e)#3}
\creflabelformat{SubFig_f}{#2#1(f)#3}
\creflabelformat{SubFig_bc}{#2#1(b-c)#3}
\creflabelformat{SubFig_cd}{#2#1(c-d)#3}
\creflabelformat{SubFig_ab}{#2#1(a-b)#3}
\creflabelformat{SubFig_ce}{#2#1(c-e)#3}
\creflabelformat{SubFig_ac}{#2#1(a-c)#3}
\creflabelformat{SubFig_df}{#2#1(d-f)#3}

\newcommand{\MyFigLabel}[1]{
  \label{#1}
  \label[SubFig_a]{#1_a}
  \label[SubFig_b]{#1_b}
  \label[SubFig_c]{#1_c}
  \label[SubFig_d]{#1_d}
  \label[SubFig_e]{#1_e}
  \label[SubFig_f]{#1_f}
  }

\newcommand{\MyFigLabelRange}[3]{\label[SubFig_#2#3]{#1_#2#3}}

\usepackage[export]{adjustbox}

\newcommand{\mytilde}{\raise.17ex\hbox{$\scriptstyle\mathtt{\sim}$}}

\def\mytitle{
	AlGaAs nanowires as a universal platform for GaAs, InGaAs, and InAs quantum dots
}
\title{\bfseries \boldmath \mytitle}

\usepackage{authblk} 
\author[1]{Rohan~Radhakrishnan}
\author[2]{Rodion~Reznik}
\author[3]{Gilles~Patriarche}
\author[2,4]{Igor~Ilkiv}
\author[2,4]{Konstantin~Kotlyar}
\author[2,4]{Anna~Andreeva}
\author[2,4]{George~Cirlin}
\author[1]{Nika~Akopian*}
\affil[1]{\small DTU Department of Electrical and Photonics Engineering, Technical University of Denmark, 2800 Kongens Lyngby, Denmark.}
\affil[2]{\small Faculty of Physics, St. Petersburg State University, Universitetskaya Embankment 7-9, 199034 St. Petersburg, Russia.}
\affil[3]{\small Centre de Nanosciences et Nanotechnologies, Université Paris Saclay, CNRS, 91120 Palaiseau, France.}
\affil[4]{\small Department of Epitaxial nanotechnologies, Alferov University, Khlopina 8/3, 194021 St. Petersburg, Russia.}

\date{\small *Corresponding author. Email: nikaak@dtu.dk}

\makeatletter
\let\MyAuthorsList\@author
\let\MyDate\@date
\makeatother

\begin{document}


\newrefsection

\maketitle


\begin{abstract} \bfseries \boldmath
Optical quantum dots (QDs) are central to photonic quantum technologies, with fabrication approaches tailored to different spectral ranges. A key challenge, however, is the realization of a unified platform—a single growth method combined with a host material offering a designable architecture—enabling wavelength tunability across the full emission range and co-integration of multiple quantum dots. Here, we introduce AlGaAs nanowires as a universal host for GaAs, InGaAs, and InAs QDs. Building on our previous demonstration of high-quality GaAs QDs, we realize InGaAs QDs with tunable emission by varying the growth duration from 2 to 5~s, achieving emission at 780 and 920~nm. We further showcase the platform's versatility for multi-quantum-dot devices by co-integrating two InGaAs QDs, as well as GaAs and InGaAs QDs within a single nanowire. Finally, we initiate a first step toward pure InAs QDs by growing a pristine InAs segment on AlGaAs nanowires, demonstrating material compatibility.
\end{abstract}


\vspace{10mm}

\noindent Quantum dots (QDs) have emerged as key building blocks for optical quantum technologies, offering coherent single-photon emission, engineered electronic states, and compatibility with integrated photonic architectures \autocite{heindelQuantumDotsPhotonic2023}. Many applications rely critically on the ability to tailor the emission wavelength of QDs across a broad spectral window, spanning from the visible to the telecom band. For example, a flying qubit to be stored in an atom-based quantum memory should match the specific atomic transition of the device, which is often in the visible/NIR \autocite{langenfeldNetworkreadyRandomaccessQubits2020a,wangEfficientQuantumMemory2019,vernaz-grisHighlyefficientQuantumMemory2018}, whereas long-distance quantum networks rely on low-loss telecom bands centered around 1.3~\textmu m (O-band) and 1.55~\textmu m (C-band) \autocite{yuTelecombandQuantumDot2023}.

The emission properties of QDs are determined primarily by their material composition and confinement environment. GaAs, InGaAs, and InAs QDs each provide access to different spectral ranges and are grown on different platforms. GaAs QDs embedded in AlGaAs emit near 780-790~nm \autocite{huberHighlyIndistinguishableStrongly2017}, while In(Ga)As QDs are usually combined with GaAs and emit in the range from near-infrared to telecom C-band depending on the growth approach \autocite{yuTelecombandQuantumDot2023}. A single platform supporting QDs across the entire range from 780~nm to C-band would eliminate the need for different host materials and simplify device fabrication by providing a standard single-photon architecture that easily adjusts to any wavelength-specific application. Moreover, such a platform would also allow co-integration of multiple quantum emitters at different wavelengths within a common architecture—crucial for future channel multiplexing \autocite{elshaariOnchipSinglePhoton2017} or entanglement-distribution schemes between wavelength-mismatched nodes through the generation of two-color entangled photon pairs \autocite{dietzFoldedsandwichPolarizationentangledTwocolor2016,pengCompactNonDegenerateEntangledPhoton}.

AlGaAs, because of its larger bandgap than GaAs and In(Ga)As, is an ideal host. However, existing fabrication approaches have intrinsic limitations: no single platform has successfully enabled the growth of multiple high-quality QDs spanning from GaAs through InGaAs to InAs compositions. For instance, the Stranski-Krastanov growth is unsuitable for GaAs QDs in AlGaAs because of the negligible lattice mismatch. Alternatively, the local droplet etching method is independent of the lattice mismatch, and GaAs QDs\autocite{huberHighlyIndistinguishableStrongly2017}, as well as InGaAs QDs\autocite{covredasilvaLowDensityInGaAsAlGaAs2026}, can be embedded in AlGaAs. Nevertheless, it has limited scalability, as the number of vertically stacked QDs is constrained by the depth of the etched nanoholes \autocite{kusterDropletEtchingDeep2016}.

Here, we address this fundamental challenge by introducing AlGaAs nanowires as a universal host platform for GaAs, InGaAs, and InAs QDs. Following our previous demonstration of high-quality GaAs QDs \autocite{leandroNanowireQuantumDots2018,leandroResonantExcitationNanowire2020}, we extend the platform to InGaAs QDs. By varying the growth time of InGaAs segments, we form QDs emitting at 780~nm and 920~nm, respectively, demonstrating direct control over the electronic and optical properties via simple growth parameters. Furthermore, we demonstrate the geometric and compositional flexibility of this platform in building versatile and scalable multi-quantum-dot devices by integrating multiple QDs into a single structure. These QDs can differ either in size—controlled by growth time—or in composition, such as GaAs and InGaAs. Finally, we take an initial step toward pure InAs QDs by growing a pristine InAs segment on AlGaAs nanowires, thereby demonstrating material compatibility and paving the way toward QDs emitting in the telecom band.

\section*{Results}

\subsection*{Growth and structural control of InGaAs QDs in AlGaAs nanowires}

\begin{figure}[htbp]
\includegraphics[width=0.8\textwidth, center]{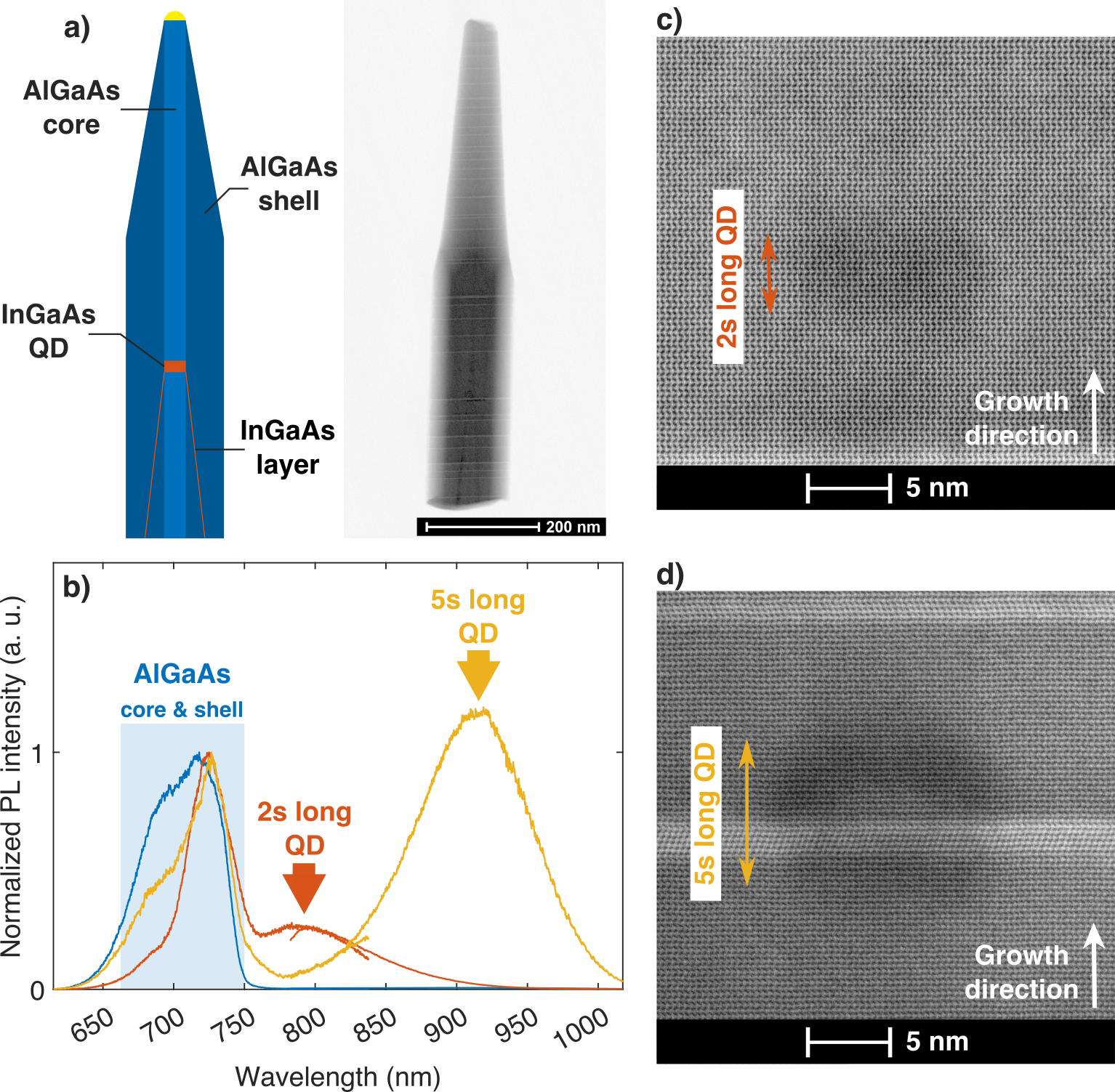}
\caption{\textbf{Structural and optical characterization of InGaAs QDs embedded in AlGaAs nanowires.} \textbf{(a)} Schematic (\textbf{left}) and low-magnification STEM image (\textbf{right}) of an AlGaAs nanowire containing an embedded InGaAs QD. The QD is formed by a short InGaAs section (2-5~s growth). The core is overgrown by an AlGaAs shell. \textbf{(b)} Normalized macro-photoluminescence spectra from nanowires containing either no QD (blue), a 2 s (orange), or 5 s (yellow) long InGaAs QDs. The QD emission can be tuned from \mytilde780 nm to \mytilde920 nm by adjusting the height of the InGaAs segment. \textbf{(c,d)} Atomic-resolution bright-field STEM images of a QD formed by a 2~s InGaAs segment \textbf{(c)}, showing no crystal-phase insertions within the QD region, and of a QD formed by a 5~s InGaAs segment \textbf{(d)}, exhibiting one crystal-phase insertion (light horizontal line) within the QD region. Both STEM images show the QD region in dark.
}
\MyFigLabel{fig:InGaAs_QD_Fig1}
\MyFigLabelRange{fig:InGaAs_QD_Fig1}{c}{d}
\end{figure}

We begin by demonstrating the controlled growth of InGaAs QDs within wurtzite AlGaAs nanowires using molecular beam epitaxy (MBE). A short interruption in AlGaAs axial growth is used to insert a defined InGaAs segment, followed by the resumption of AlGaAs deposition. \Cref{fig:InGaAs_QD_Fig1_a} shows a schematic and low-magnification scanning transmission electron microscopy (STEM) image of a nanowire containing an embedded InGaAs QD. The nanowires grow naturally into a core-shell structure with a lower Al content in the core than in the shell \autocite{dubrovskiiOriginSpontaneousCore2016a,leandroWurtziteAlGaAsNanowires2020}. We note the short zincblende insertions (visible as light and fine horizontal lines) that randomly appear along the wurtzite nanowire, as well as the presence of a thin InGaAs layer that originates from lateral growth. The InGaAs layer does not emit (see \Cref{InGaAs - SI:Absence of emission from the thin InGaAs layer}), which we attribute to its thickness being too thin to localize charge carriers\autocite{lambkinThermalQuenchingPhotoluminescence1990}. The In\textsubscript{x}Ga\textsubscript{1-x}As QD has an In content x of \mytilde0.40 (measured by energy-dispersive X-ray spectroscopy (EDX), see \Cref{InGaAs_QD - SI:EDX}) and a diameter of \mytilde10~nm, defined by the nanowire core, which is set by the gold catalyst droplet diameter. By varying the duration of InGaAs deposition between 2~s and 5~s, we deterministically tune the QD height, and correspondingly the emission wavelength.

The macro-photoluminescence spectra in \Cref{fig:InGaAs_QD_Fig1_b}, where emission is collected over numerous nanowires, confirm this tunability, with QD emission shifting from \mytilde780~nm (in red) to \mytilde920~nm (in yellow) depending on dot height. The nominal Al/(Ga+As) ratio for the three samples is set to 0.4 to tune the emission from the AlGaAs nanowire to 650 - 750~nm\autocite{leandroWurtziteAlGaAsNanowires2020} and therefore avoid spectral overlap with the QD emission, as confirmed by the photoluminescence of nanowires without QDs (in blue).

Atomic-resolution STEM images (\Cref{fig:InGaAs_QD_Fig1_cd}) show that increasing the InGaAs growth time from 2 to 5~s increases the QD height from \mytilde5 to \mytilde10~nm. They also reveal short zincblende insertions within the wurtzite InGaAs segment, which we attribute to a phase instability induced by In/Al flux commutation. While examples of pure wurtzite 2~s QDs were possible to find, because they are short enough to be free of any zincblende insertions, this was not the case for the 5~s QDs. To understand the impact of these crystal-phase insertions on the QD emission, we measured photoluminescence spectra of nanowires transferred onto a transmission electron microscopy grid, thereby establishing a one-to-one correspondence between emission and crystal structure (see \Cref{InGaAs_QD - SI:Effect of crystal-phase insertions on the InGaAs QDs}). We conclude that the crystal-phase insertions modulate the QD emission wavelength \autocite{francavigliaTuningAdatomMobility2018}, but do not split the QD into multiple distinct ones emitting at different spectral ranges.

These results establish that we can reliably tune the structural and optical properties of InGaAs QDs by controlling the segment length within the AlGaAs nanowire host.

\subsection*{Excitonic complexes and single-photon emission}

\begin{figure}[htbp]
\includegraphics[width=0.8\textwidth, center]{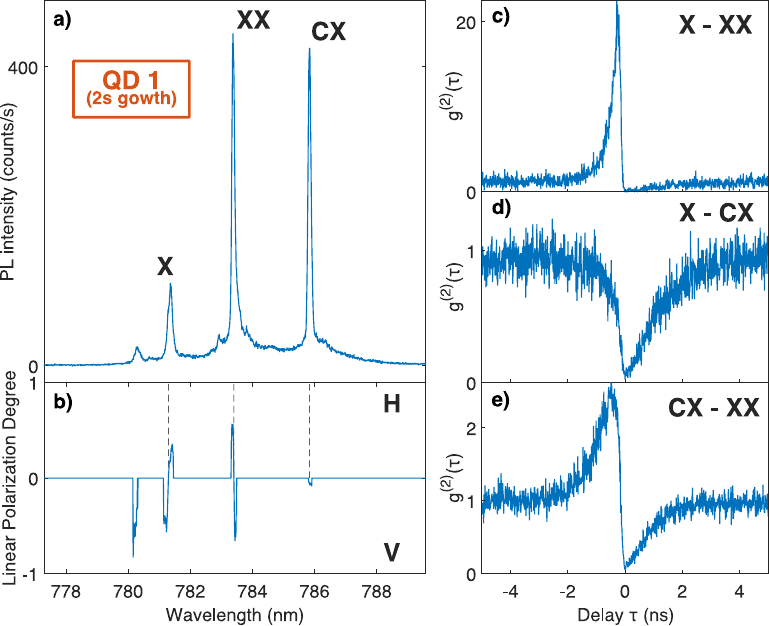}
\caption{\textbf{Optical characterization of a 2~s InGaAs QD embedded in an AlGaAs nanowire.} \textbf{(a)} Micro-photoluminescence spectrum of a single QD (QD~1) showing neutral exciton (X), biexciton (XX), and charged exciton (CX) transitions. \textbf{(b)} Degree of linear polarization for the same dot, with clear switching between horizontal (H) and vertical (V) polarization among the spectral lines, showing fine-structure splitting of the X state. Noise has been removed from the spectra used to calculate the degree of linear polarization to avoid unrealistic spikes. \Cref{InGaAs_QD - SI:Polarization-dependent PL of 2s QD} presents the full polarization-dependent photoluminescence. \textbf{(c - e)} Cross-correlation measurements between X-XX \textbf{(c)}, X-CX \textbf{(d)}, and CX-XX \textbf{(e)} transitions, showing pronounced bunching and antibunching behavior consistent with cascaded emission dynamics and single-photon statistics. The order in the state labeling for each panel indicates the start and stop channel for positive delay, and the opposite order is valid for negative delay.
} 
\MyFigLabel{fig:InGaAs_QD_Fig2}
\MyFigLabelRange{fig:InGaAs_QD_Fig2}{c}{e}
\end{figure}

Next, we examine the optical properties of individual QDs. \Cref{fig:InGaAs_QD_Fig2} shows one example of a 2~s InGaAs QD (QD~1). The spectrum (\Cref{fig:InGaAs_QD_Fig2_a}) reveals distinct exciton (X), biexciton (XX), and charged exciton (CX) transitions. Linear polarization measurements (\Cref{fig:InGaAs_QD_Fig2_b}) show alternating polarization orientations, reflecting fine-structure splitting of the exciton state due to electron-hole anisotropic confinement\autocite{bayerFineStructureNeutral2002}. Cross-correlation measurements between each emission line (\Cref{fig:InGaAs_QD_Fig2_ce}) exhibit characteristic bunching and antibunching features, confirming the typical dynamics of such a multi-level system. We observe a strong bunching effect between the biexciton and the exciton (\Cref{fig:InGaAs_QD_Fig2_c}), indicating cascaded emission. The bunching is much lower between the biexciton and the charged exciton (\Cref{fig:InGaAs_QD_Fig2_e}), due to the additional charging time in the process. The charging/discharging times also induce the asymmetry observed in the correlation between the exciton and charged exciton (\Cref{fig:InGaAs_QD_Fig2_d}). We measure for all three cross-correlation measurements, along with the autocorrelation measurements (see \Cref{InGaAs_QD - SI:Further optical characterization of the 2~s InGaAs QD}), a g\textsuperscript{2}(0) close to zero, confirming single-photon emission. Additionally, the observed nanosecond timescales are consistent with the lifetime of \mytilde1~ns measured for the exciton and charged exciton (see \Cref{InGaAs_QD - SI:Further optical characterization of the 2~s InGaAs QD}, along with further optical characterizations of QD~1).

\begin{figure}[htbp]
\includegraphics[width=0.8\textwidth, center]{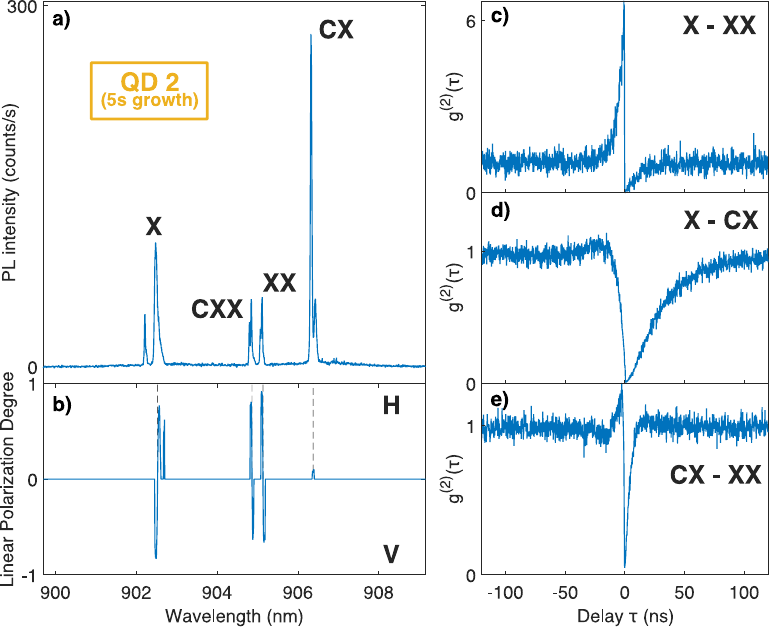}
\caption{\textbf{Optical characterization of a 5~s InGaAs QD embedded in an AlGaAs nanowire.} \textbf{(a)} Micro-photoluminescence spectrum of a single QD (QD~2) showing neutral exciton (X), biexciton (XX), charged exciton (CX), and charged biexciton (CXX) transitions. \textbf{(b)} Degree of linear polarization for the same dot, with clear switching between horizontal (H) and vertical (V) polarization among the spectral lines, showing fine-structure splitting of the X state. Noise has been removed from the spectra used to calculate the degree of linear polarization to avoid unrealistic spikes. \Cref{InGaAs_QD - Polarization-dependent PL of 5s QD} presents the full polarization-dependent photoluminescence. \textbf{(c-e)} Cross-correlation measurements between X-XX \textbf{(c)}, X-CX \textbf{(d)}, and CX-XX \textbf{(e)} transitions, showing pronounced bunching or antibunching behavior consistent with cascaded emission dynamics and single photon statistics. The order in the state labeling for each panel indicates the start and stop channel for positive delay, and the opposite order is valid for negative delay.
}
\MyFigLabel{fig:InGaAs_QD_Fig3}
\MyFigLabelRange{fig:InGaAs_QD_Fig3}{c}{e}
\end{figure}

Similar behavior is observed for 5~s QDs, and \Cref{fig:InGaAs_QD_Fig3} presents one example (QD~2). Here, the photoluminescence spectrum (\Cref{fig:InGaAs_QD_Fig3_a}) includes an additional transition assigned to a charged biexciton (CXX) state \autocite{poemRadiativeCascadesCharged2010,cadeFineStructureMagnetooptics2006,kettlerNeutralChargedBiexcitonexciton2016}. Polarization analysis (\Cref{fig:InGaAs_QD_Fig3_b}) again reveals alternating polarization axes for the exciton and mirrored behavior for the biexciton and charged biexciton. Cross-correlation measurements (\Cref{fig:InGaAs_QD_Fig3_ce}) confirm cascaded emission dynamics and charging/discharging processes. We note longer timescales than in the 2~s quantum dot case (QD 1), indicating the presence of a slow process in the multi-level dynamics; identifying its origin would require further study. We also found 5~s QDs with faster dynamics similar to the 2~s QD case; one example is presented in \Cref{InGaAs_QD - Further optical characterization of the 5~s InGaAs QD}, together with a more detailed optical study of QD~2.

Together, these measurements demonstrate the high optical quality of our QDs across different emission energies.

\subsection*{A universal and scalable platform for multi-quantum-dot architectures}

We now showcase the geometrical and compositional versatility of our platform for building a multi-quantum-dot structure by integrating two QDs within a single nanowire. \Cref{fig:InGaAs_QD_Fig4_ab} show the photoluminescence spectra of two individual nanowires grown with different designs.

In the first design, a 2~s and a 5~s InGaAs QD are embedded, and emit at 780~nm and 900~nm, respectively (\Cref{fig:InGaAs_QD_Fig4_a}), consistent with results on single-quantum-dot nanowires (\Cref{fig:InGaAs_QD_Fig1_b}). The bottom-up nanowire architecture enables straightforward growth of a second QD by simply adding a second InGaAs growth cycle. We first grow a 2~s QD and then a 5~s QD, with a target spacing of 5~nm to keep them close while sufficiently separated to suppress directional phonon-assisted tunneling\autocite{khoshnegarSolidStateSource2017a,reischleNonresonantTunnelingSingle2007}. Observing the two QD emissions spectrally distinct confirms that they remain physically separated despite their close spacing, highlighting the sharp AlGaAs/InGaAs interfaces required for scalable structures with tightly spaced QDs\autocite{liLocationQubitsMultipleQuantumDot2024a}.

The second design demonstrates compositional flexibility by integrating two QDs grown with different materials into the same nanowire: a 5~s InGaAs QD with a GaAs QD, formed by turning off the Al flux for 17~s\autocite{cirlinAlGaAsAlGaAsGaAs2017a}. The growth time of the GaAs QD is chosen to target emission at 780~nm\autocite{leandroResonantExcitationNanowire2020}, and the two QDs are here separated by \mytilde2~\textmu m. Micro-photoluminescence (\Cref{fig:InGaAs_QD_Fig4_b}) clearly resolves both emitters, with the GaAs QD at 780~nm and the InGaAs QD at 850~nm, which matches the previously obtained spectral range by macro-photoluminescence on single 5~s QDs in AlGaAs nanowires (\Cref{fig:InGaAs_QD_Fig1_b}).

\begin{figure}[htbp]
\includegraphics[width=0.8\textwidth, center]{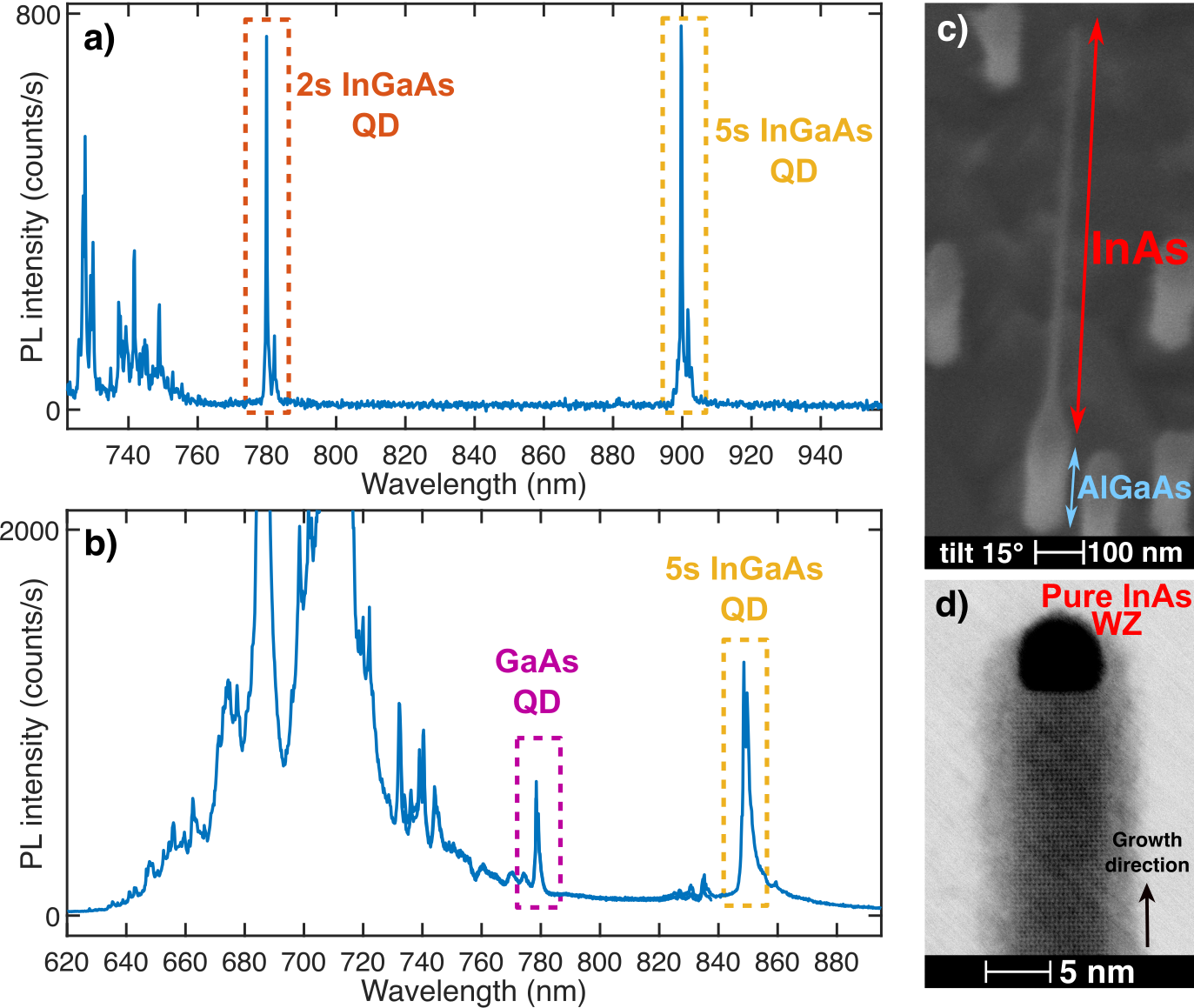}
\caption{\textbf{Two different QDs in a nanowire, and a pure InAs segment on top of a nanowire.} \textbf{(a)} Micro-photoluminescence spectrum of a 2~s and 5~s InGaAs QDs embedded in the same nanowire. \textbf{(b)} Micro-photoluminescence spectrum of a GaAs and 5~s InGaAs QDs embedded in the same nanowire. \textbf{(c)} Scanning electron microscopy (SEM) image of an InAs segment on top of an AlGaAs nanowire. \textbf{(d)} Atomic-resolution bright-field STEM of an InAs segment showing pure wurtzite crystal phase without any crystal-phase insertions  (an enlargement is available in \Cref{InGaAs - SI:Compositional analysis of the InAs segments grown on top of the AlGaAs nanowires}.)}
\MyFigLabel{fig:InGaAs_QD_Fig4}
\MyFigLabelRange{fig:InGaAs_QD_Fig4}{a}{b}
\MyFigLabelRange{fig:InGaAs_QD_Fig4}{c}{d}
\end{figure}

\subsection*{Extending the platform to InAs segments}

Finally, to complete the material span from GaAs to InAs within a single host system, we take a first step toward InAs QDs by demonstrating, for the first time, the growth of InAs segments on top of AlGaAs nanowires (\Cref{fig:InGaAs_QD_Fig4_c}), establishing material compatibility between InAs and AlGaAs. Atomic-resolution STEM of the InAs segment shows a pure wurtzite phase (\Cref{fig:InGaAs_QD_Fig4_d}), and EDX measurements confirmed its compositional purity (see \Cref{InGaAs - SI:Compositional analysis of the InAs segments grown on top of the AlGaAs nanowires}).

These results provide a key foundation for extending our platform toward QDs in the telecom regime. Taken together with our InGaAs and GaAs results \autocite{leandroNanowireQuantumDots2018,leandroResonantExcitationNanowire2020}, this establishes AlGaAs nanowires as the first and, to our knowledge, only platform supporting QD growth across the full GaAs-InAs compositional range.

\section*{Conclusion and outlook}

We have introduced AlGaAs nanowires as a versatile platform for high-quality InGaAs QDs. Individual QDs exhibit clean excitonic transitions with narrow linewidths, fine-structure splitting, single-photon statistics, and cascaded dynamics. The emission wavelength is controllably tuned through segment height, offering a direct route to target specific spectral regions. Moreover, we demonstrated the integration of two QDs with different sizes and materials into a single nanowire without fundamentally modifying the growth process. It showcases the high designability and flexibility of the bottom-up nanowire architecture for building multi-quantum-dot structures, with the added advantage of AlGaAs as a host material, enabling the growth of QDs at arbitrary compositions from GaAs to InAs using a single, unified protocol.

This work lays the foundation for a new class of tunable, application-adaptable quantum light sources. AlGaAs nanowires emerge not only as a structurally and optically robust host but also as a unifying material platform for hybrid quantum networks that require tunable sources across a wide spectral range—from near-infrared to telecom wavelengths. This approach can enable on-chip architectures combining distinct quantum emitters for multi-wavelength operation and spectral multiplexing\autocite{elshaariOnchipSinglePhoton2017,laferriereMultiplexedSinglePhotonSource2020,wengerowskyEntanglementbasedWavelengthmultiplexedQuantum2018}. Furthermore, locally entangling multiple embedded QDs via neighbor interactions\autocite{liLocationQubitsMultipleQuantumDot2024a,khoshnegarSolidStateSource2017a,hacklExperimentalProposalProbe2023} would create a unique quantum repeater capable of interconnecting wavelength-mismatched nodes over a wide spectral range\autocite{dietzFoldedsandwichPolarizationentangledTwocolor2016,pengCompactNonDegenerateEntangledPhoton}—a key step toward scalable and spectrally flexible quantum networks.

\section*{Methods}

\paragraph{Nanowire growth}
The nanowires were grown on Si(111) substrates in an MBE machine equipped with effusion cells for Ga, Al, In, and As\textsubscript{4}, as well as a separate metallization chamber for Au deposition. Before growth, the oxide on the substrate surface is removed by wet chemical treatment in an HF:H\textsubscript{2}O solution, followed by annealing at 850°C in the metallization chamber. Afterward, the temperature was reduced to 550°C, and we deposited a \mytilde0.5~nm-thick Au layer that we let rest for 1~min to enhance droplets homogeneity. The substrate was then transferred to the growth chamber without breaking the vacuum. After reaching the growth temperature (510°C) and stabilization of the As\textsubscript{4} flux, the Al and Ga sources are simultaneously opened to initiate the growth. 

For the three samples with 2~s and 5~s InGaAs QDs and the reference sample without any insertion, the total growth time was 25~min, and the QDs were grown after 12~min by switching between the Al and In fluxes for a short time (2~s and 5~s). The nominal Al and In contents were respectively $\mathrm{x_{Al} = 0.4}$ and $\mathrm{x_{In} = 0.5}$. The material fluxes were calibrated using a separate GaAs(100) substrate and corresponded to the growth rates of planar layers: GaAs - 0.6~monolayers per second (ML/s), AlAs - 0.4~ML/s, and InAs - 0.6~ML/s. All samples were grown under As-rich conditions.

The nanowires with an InGaAs and a GaAs QD were grown using a similar recipe, and a short interruption of the Al flux formed the GaAs QD. We used the following sequence: 30~min of AlGaAs, 5~s of InGaAs, 60~min of AlGaAs, 17~s of GaAs, and finally 20~min of AlGaAs. The Al and In content were the same as previously.

For the nanowire with a long InAs segment, we stopped the AlGaAs growth after 15~min, reduced the substrate temperature to 270°C, and then grew the InAs for 10~min with a growth rate of \mytilde0.25 ML/s for the planar layer. The nominal Al content for this sample is $\mathrm{x_{Al} = 0.3}$.

\paragraph{Transmission electron microscopy and EDX}
The samples were mostly observed in an aberration-corrected FEI TITAN 200 TEM/STEM microscope or an aberration-corrected Jeol2200FS TEM/STEM microscope, both operating at 200~keV. On the FEI Titan microscope in STEM mode, the convergence half-angle of the probe was 17.6~mrad and the detection inner and outer half-angles for high-angle annular dark-field STEM (HAADF-STEM) were 69~mrad and 200~mrad, respectively.

All STEM images presented in the main text were acquired using bright field (BF-STEM), which is sensitive to the deformation field induced by the InGaAs QD, due to its lattice mismatch with the surrounding AlGaAs matrix.

The nanowires were dispersed on a carbon membrane (placed on a copper grid) and then oriented along the <11-20> zone axis of the wurtzite structure (corresponding to the <110> zone axis of the cubic zinc blende structure), the only orientation allowing the two possible cubic and hexagonal crystalline phases to be distinguished.

All micrographs were 2048 by 2048 pixels. The dwell time was 8~\textmu s and the total acquisition time 41~s. Additional EDX measurements were performed in the Titan microscope featuring the Chemistem system, that uses a Bruker windowless Super-X four-quadrant detector and has a collection angle of 0.8~sr. During EDX acquisition, the sample was also aligned along <11-20>. The acquisition time for the linescans was 10~min, during which no significant drift occurred. EDX chemical maps are obtained with acquisition times between 20 and 40 minutes, with drift correction via cross-correlation The quantification of EDX analysis profiles is obtained using the Cliff-Lorimer method with correction for absorption effects for light elements such as aluminum. Reference standards have been established to refine the k-factors used for quantification \autocite{pantzasExperimentalQuantificationAtomicallyresolved2021}. 

\paragraph{Photoluminescence and polarization-resolved measurements}
Photoluminescence measurements were performed at a low temperature (\mytilde6~K), using a 532~nm green continuous wave laser and a 750~mm focal length spectrometer equipped with a low-noise Peltier-cooled charge-coupled device (CCD). The entrance slit of the spectrometer was open to 30~\textmu m, and the CCD has a pixel size of 20~\textmu m. Macro-photoluminescence was measured with a 5x objective with a low NA of 0.12, and the spectra were acquired over 10~s at the same time as scanning the sample with a piezoelectric stage to average over numerous nanowires. Microphotoluminescence measurements are performed on single nanowires using a 100x 0.9NA objective. The high-resolution spectra from \Cref{fig:InGaAs_QD_Fig2,fig:InGaAs_QD_Fig3}, and all the polarization-resolved photoluminescence measurements are taken with a 1800~gr/mm holographic grating providing a spectral resolution of \mytilde0.02~nm. The spectra presented in \Cref{fig:InGaAs_QD_Fig1_b} and \Cref{fig:InGaAs_QD_Fig4_b} have been realized by stitching two spectra acquired with a 150~gr/mm ruled grating tilted for different central wavelengths. The same grating has been used for \Cref{fig:InGaAs_QD_Fig4_a}. For the polarization-resolved photoluminescence measurements, a half-wave plate placed in front of a polarizing beam splitter was rotated to control the polarization. A second half-wave plate was placed in front of the spectrometer to account for the preferential polarization of the grating and optimize the count efficiency.

\paragraph{Photon correlation and time-resolved photoluminescence measurements}
Second-order correlation measurements were performed under continuous-wave excitation (532~nm laser) and using superconducting nanowire single-photon detectors (SNSPDs) connected to a time tagger. The detector timing jitter (FWHM) is 30~ps, while it is 72~ps for a single channel of the time tagger. For cross-correlation, the two specific emission lines were separately filtered out of the full spectrum with the use of two tunable reflective Bragg filters with an ultra-narrow linewidth of \mytilde0.3~nm. The light was then coupled to a single-mode fiber. In the case of autocorrelation, a single filter was used, along with a fiber-based beam splitter to form a Hanbury Brown and Twiss (HBT) interferometer. Time-resolved photoluminescence measurements were performed with a 440 nm diode pulsed laser, and a 2.5 to 20~MHz repetition frequency was used depending on the lifetime value. The FWHM of the pulses was measured to be 200~ps.


\section*{Acknowledgments}

We thank Akaash Srikanth and Mikaela Ppasiou for their help in measuring the photoluminescence of nanowires embedding two InGaAs QDs. N. Akopian gratefully acknowledges the financial support from the European Research Council (ERC Grant Agreement No. 101003378) and from the Carlsberg Foundation (grant No. CF22-1543). G. Patriarche thanks the French technology network Renatech for its participation in the financial support of the C2N characterization platform, used to perform TEM and EDX measurements. Growth experiments were carried out with the support of the Russian Science Foundation (grant No. 25-79-10101).

\section*{Author contributions}

R. Radhakrishnan, R. Reznik, G. Cirlin, and N. Akopian conceived the original concept and designed the experiment. R. Reznik, I. Ilkiv, K. Kotlyar, A. Andreeva, and G. Cirlin grew samples and performed SEM characterization. G. Patriarche provided the STEM data and structural analysis. R. Radhakrishnan conducted the optical experiments, which included building the setup and performing measurements. R. Radhakrishnan and N. Akopian discussed the results and wrote the manuscript with inputs from all authors. N. Akopian coordinated the project.


\printbibliography


\newpage


\renewcommand{\thefigure}{S\arabic{figure}}
\renewcommand{\thetable}{S\arabic{table}}
\renewcommand{\theequation}{S\arabic{equation}}
\renewcommand{\thepage}{S\arabic{page}}
\setcounter{figure}{0}
\setcounter{table}{0}
\setcounter{equation}{0}
\setcounter{section}{0}
\setcounter{page}{1} 

\begin{center}
\newrefsection 

\section*{Supplementary Materials for\\ \mytitle}

{\large
\MyAuthorsList
}\\
\MyDate
\end{center}

\subsubsection*{This PDF file includes:}
Supplementary Text\\
Figures S1 to S12\\

\newpage
{\hypersetup{linkcolor=black}
\tableofcontents}

\newpage

\section{Compositional analysis of the InGaAs QDs}
\label[SI]{InGaAs_QD - SI:EDX}

\begin{figure}[htbp]
\includegraphics[width=\textwidth, center]{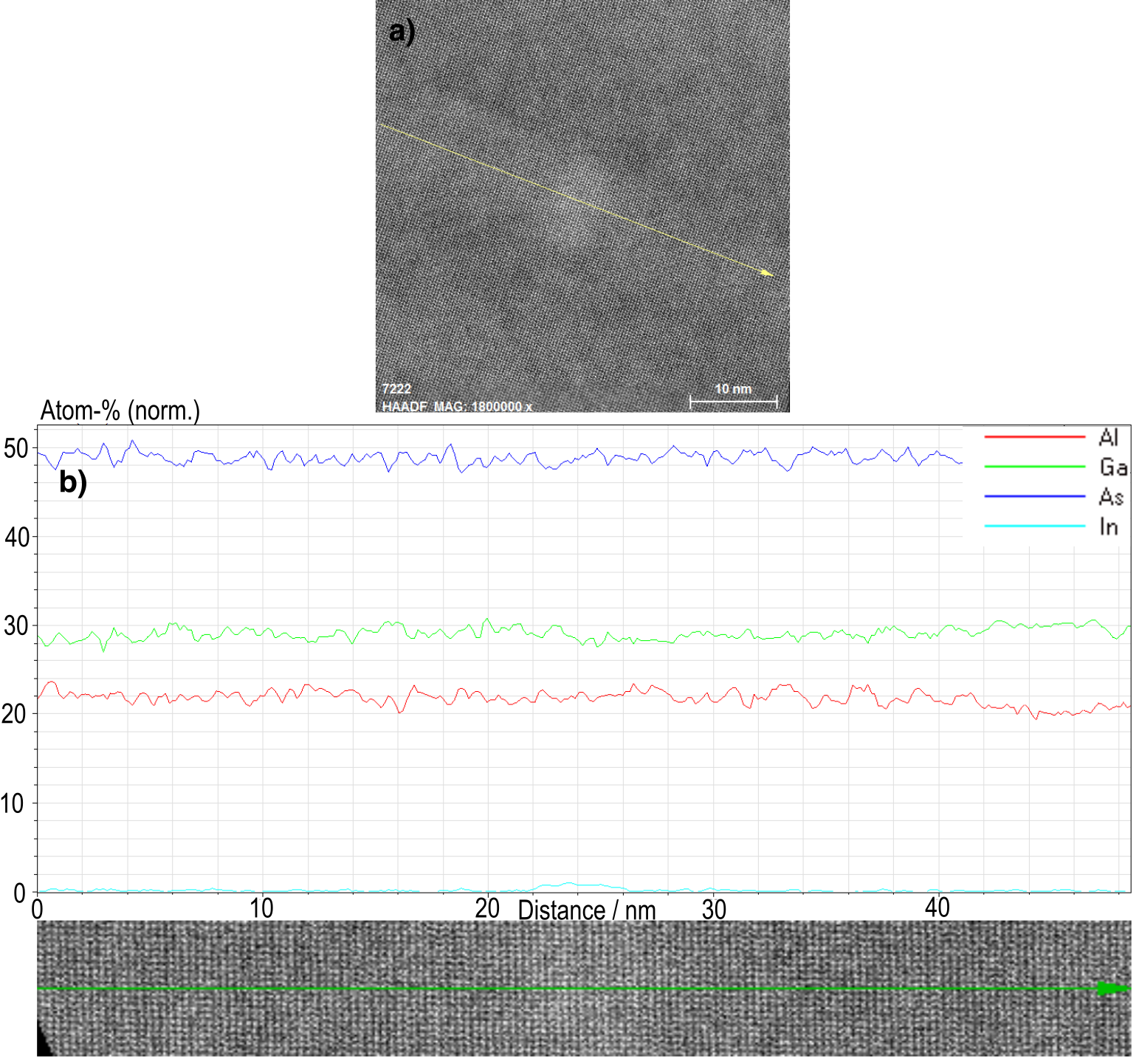}
\caption{\textbf{EDX linescan of a 2~s-long QD.} High-angle annular dark-field (HAADF) STEM image \textbf{(a)} and corresponding EDX linescan \textbf{(b)} across the InGaAs segment of a 2~s long QD. The arrows indicate the line and direction of the scan. We detect a small indium content (weak signal below 2\%) at the position of the QD; the low value is due to the screening of the thick AlGaAs shell.}
\MyFigLabel{fig:InGaAs_QD_EDX_InGaAs}
\end{figure}

To confirm the presence of indium in the quantum dots (QDs), we analyze the chemical composition of the segment through energy-dispersive X-ray spectroscopy (EDX) for seven nanowires with a 2~s InGaAs QD and five nanowires with a 5~s InGaAs QD. \Cref{fig:InGaAs_QD_EDX_InGaAs} presents one example. All of them show the presence of indium at the QD position, despite the thick shell that screens and drastically reduces the signal. Indeed, the nanowire diameter is \mytilde150~nm, 15 times larger than the core diameter (\mytilde10~nm), which prevents us from measuring the actual indium content in the QD. To overcome this, we grow a sample with the 5~s QD at the top of the nanowire, where the shell is the thinnest, and measure the QD composition by EDX on eight nanowires. We find an average In/(In+Ga) ratio of 0.40, with a large distribution (standard deviation = 0.13). 

\section{Further optical characterization of the 2~s InGaAs QD}
\label[SI]{InGaAs_QD - SI:Further optical characterization of the 2~s InGaAs QD}

We present in this section additional optical characterization of the 2~s InGaAs QD studied in the main text (QD~1).

    \subsection{Power series}

    \begin{figure}[htbp]
    \includegraphics[width=\textwidth, center]{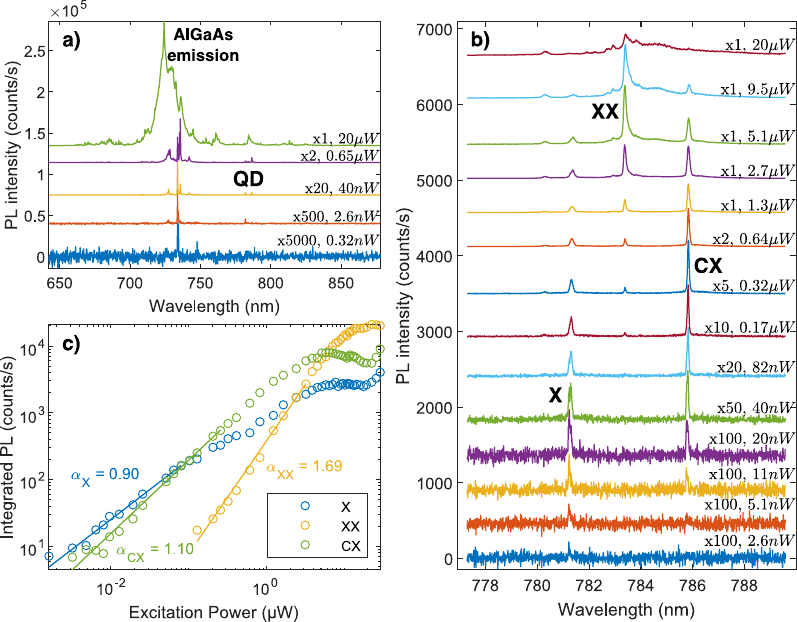}
    \caption{\textbf{Power-dependent photoluminescence of the 2~s InGaAs QD.} \textbf{(a)} Low-resolution spectroscopy of the full nanowire emission spectrum measured at different powers and with a 150~gr/mm grating. The graphs are staggered to increase visibility. \textbf{(b)} High-resolution spectroscopy of the QD emission at different powers and measured with a 1800~gr/mm grating. \textbf{(c)} Integrated number of counts over the FHWM for each emission line (X, XX, and CX) as a function of the excitation power. The intensity $I$ below saturation is fitted with the power law $I \propto  Power^{\alpha}$. The intensity is extracted from spectra acquired with a 1200~gr/mm grating.}
    \MyFigLabel{fig:InGaAs_QD_2s_PPL}
    \MyFigLabelRange{fig:InGaAs_QD_2s_PPL}{b}{c}
    \end{figure}

    The low-resolution photoluminescence spectra from \Cref{fig:InGaAs_QD_2s_PPL_a} acquired at different excitation powers show the QD emission at 785~nm and the AlGaAs core-shell emission below 760~nm. We can notice that, despite the InGaAs QD being the lower potential region, the power series begins with AlGaAs emission. This behavior has also been observed in other nanowires and is explained by carriers being trapped in the crystal-phase QDs present in the AlGaAs core, which prevents them from diffusing toward the InGaAs QD.

    The evolution with the excitation power of the different emission lines follows well our attribution to QD states (\Cref{fig:InGaAs_QD_2s_PPL_bc}). The neutral exciton line (X) is the first one to appear at low power, followed by the charged exciton line (CX). They both have close to linear power dependency, and they saturate and then decrease when the biexciton line (XX) becomes predominant. As expected, the biexciton evolution is closer to a quadratic power dependency \autocite{finleyChargedNeutralExciton2001}. The divergence from the ideal case $\alpha_{X} = 1$ for the exciton and $\alpha_{XX} = 2$ for the biexciton, could be explained by a power dependency of the charging rate of the QD: the probability of having an additional charge in the QD increases with excitation power, which slows down the power evolution of the neutral states ($\alpha_{X} < 1$ and $\alpha_{XX} < 2$) and speeds up the charged state's one ($\alpha_{CX} > 1$)).

    \subsection{Polarization-dependent photoluminescence and degree of linear polarization}
    \label[SI]{InGaAs_QD - SI:Polarization-dependent PL of 2s QD}

    \begin{figure}[htbp]
    \includegraphics[width=\textwidth, center]{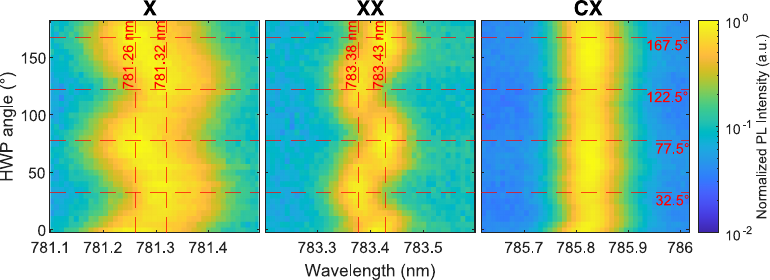}
    \caption{\textbf{Polarization-dependent photoluminescence of the 2~s InGaAs QD.} The exciton (X) and biexciton (XX) lines show a fine-structure splitting of $110 \pm 30 ~\mu eV$, while the charged exciton wavelength stays constant with polarization.}
    \MyFigLabel{fig:InGaAs_QD_2s_Pola}
    \end{figure}
    
    We measure the polarization dependency of the photoluminescence by placing a half-wave plate before the spectrometer, followed by a polarizing beam splitter that acts as a linear polarizer. We then rotate the wave plate by an angle $\theta$ to rotate the measured polarization by an angle $2\theta$. Orthogonal polarizations are then separated by 45\textdegree.

    The emission lines' labeling is confirmed by the presence of fine-structure splitting on the exciton state (\Cref{fig:InGaAs_QD_2s_Pola}), due to the electron-hole anisotropic confinement of the QD. Because of the cascade emission, the biexciton line also shows the same fine-structure splitting with a mirrored polarization behavior. On the other hand, the charge exciton intensity remains constant due to its circular polarization. We calculate the degree of linear polarization (DLP) presented in the main text with the following formula:

    \begin{align}
        \label{eq:InGaAs_QD_DLP}
        DLP = \frac{I_H - I_V}{I_H + I_V}
    \end{align}

    where $I_H$ ($I_V$) is the photoluminescence intensity when the half-wave plate angle is set to 32.5\textdegree~(77.5\textdegree). We also set a noise threshold to remove unrealistic spikes originating from noise fluctuations having a value for $I_H + I_V$ close to zero.

    \subsection{Autocorrelation and lifetime measurements}

    \begin{figure}[htbp]
    \includegraphics[width=0.8\textwidth, center]{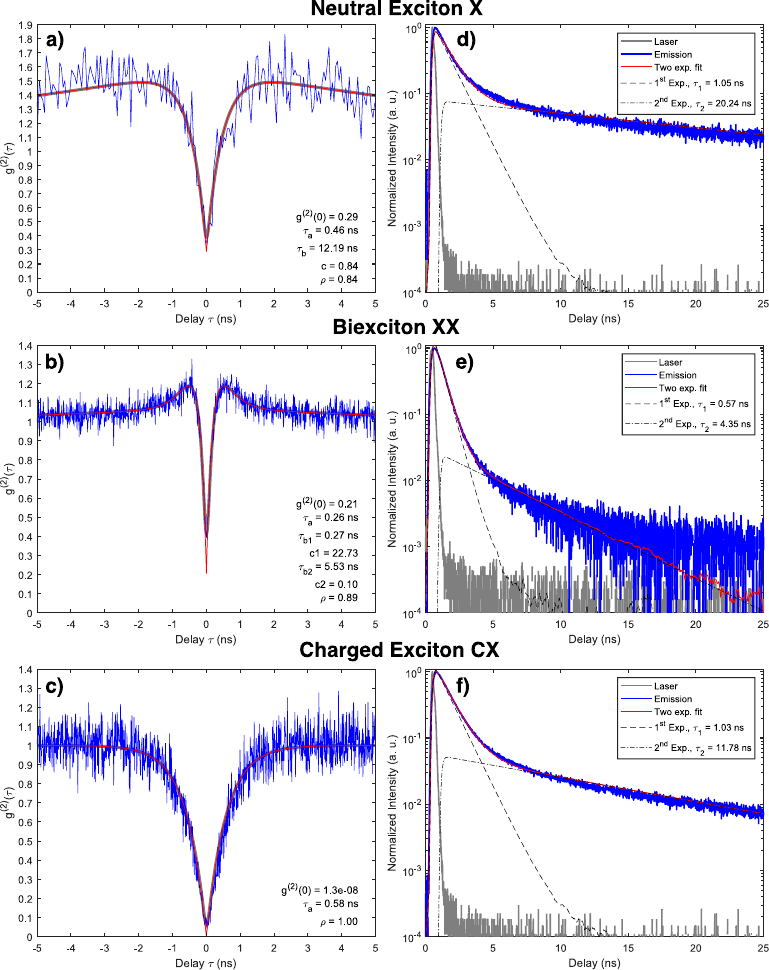}
    \caption{\textbf{Photon statistics and lifetime measurements of the 2~s InGaAs QD.} \textbf{(a - c)} Autocorrelation measurements of the exciton, biexciton, and charged exciton. The red line is the result of the fit (transparent gray line) of the g\textsuperscript{(2)} function deconvoluted by the instrument response function of the measuring setup (Gaussian with a FWHM of 110~ps). \textbf{(d - f)} Time-resolved photoluminescence showing the lifetime of the different states. To extract the lifetime, the data are fitted with decaying exponential functions convoluted with the laser pulse shape (gray line).}
    \MyFigLabel{fig:InGaAs_QD_2s_Corr}
    \MyFigLabelRange{fig:InGaAs_QD_2s_Corr}{a}{c}
    \MyFigLabelRange{fig:InGaAs_QD_2s_Corr}{d}{f}
    \end{figure}

    In addition to the cross-correlation measurements presented in the main text, we also measure autocorrelation (\Cref{fig:InGaAs_QD_2s_Corr_ac}). For each line, the value for g\textsuperscript{(2)}(0) is below 0.5, which confirms the single-photon nature of the emission. Furthermore, to extract the characteristic times, we fit the data and use different g\textsuperscript{(2)} functions for each measurement. For the charged exciton CX, we use the function:

    \begin{align}
        \label{eq:InGaAs_QD_g2_1}
        g^{(2)}(\tau) = 1 - \rho^2 \cdot e^{-\frac{|\tau - \tau_0|}{\tau_a}}
    \end{align}
 
    where $\tau_a$ is the antibunching time, $\rho$ is the signal-to-noise ratio, and $\tau_0$ is a fixed delay between the two paths of the Hanbury Brown and Twiss interferometer. The x-axis has been shifted to set the origin to $\tau_0$, making this value not visible on the figure. 

    For the exciton X, on the other hand, a weak and long bunching is observed on the shoulders of the dip, and it indicates a higher probability than average of having a second neutral exciton populating the dot just after the first one recombines and leaves the dot empty \autocite{regelmanSemiconductorQuantumDot2001}. We take it into account and use the following function:

    \begin{align}
        \label{eq:InGaAs_QD_g2_2}
        g^{(2)}(\tau) = 1 + \rho^2 \left( c \cdot e^{-\frac{|\tau - \tau_0|}{\tau_b}} - (1 + c) \cdot e^{-\frac{|\tau - \tau_0|}{\tau_a}} \right)
    \end{align}

    where $\tau_b$ is the bunching time and $c$ is the bunching amplitude. 

    The coincidence counts for the biexciton could not be fitted with the previous formula because of the presence of an additional bunching, which is much shorter in duration. We add then a second bunching term to the fitting function:

    $$ g^{(2)}(\tau) = 1 + \rho^2 \left( c_1 \cdot e^{-\frac{|\tau - \tau_0|}{\tau_{b1}}} + c_2 \cdot e^{-\frac{|\tau - \tau_0|}{\tau_{b2}}} - (1 + c_1 + c_2) \cdot e^{-\frac{|\tau - \tau_0|}{\tau_a}} \right) $$

    where $\tau_{b1}$ and $c_1$ are the parameters for the short bunching, and $\tau_{b1}$ and $c_1$ are the ones for the long bunching. 
    
    Identifying the corresponding reason for each bunching would require further study of the system through rate equations. However, we can comment on the antibunching time, which is around \mytilde0.5~ns for the exciton and the charged exciton, and is twice shorter for the biexciton ($\tau_a = 0.26~ns$). We observe the same trend for the lifetimes measured at short delay: \mytilde1~ns for X and CX, and 0.57~ns for the XX (\Cref{fig:InGaAs_QD_2s_Corr_df}). It is a sign of the two decaying paths of the biexciton, i.e., either one of the two excitons recombines, which divides the biexciton lifetime by two compared to the exciton lifetime. The presence of an additional long decay component for each lifetime can be explained by the recapturing of charge carriers originating from a long-standing reservoir in the vicinity.

\section{Further optical characterization of the 5~s InGaAs QD}
\label[SI]{InGaAs_QD - Further optical characterization of the 5~s InGaAs QD}

We present in this section additional optical characterization of the 5~s InGaAs QD studied in the main text (QD~2).

    \subsection{Power series}

    \begin{figure}[htbp]
    \includegraphics[width=\textwidth, center]{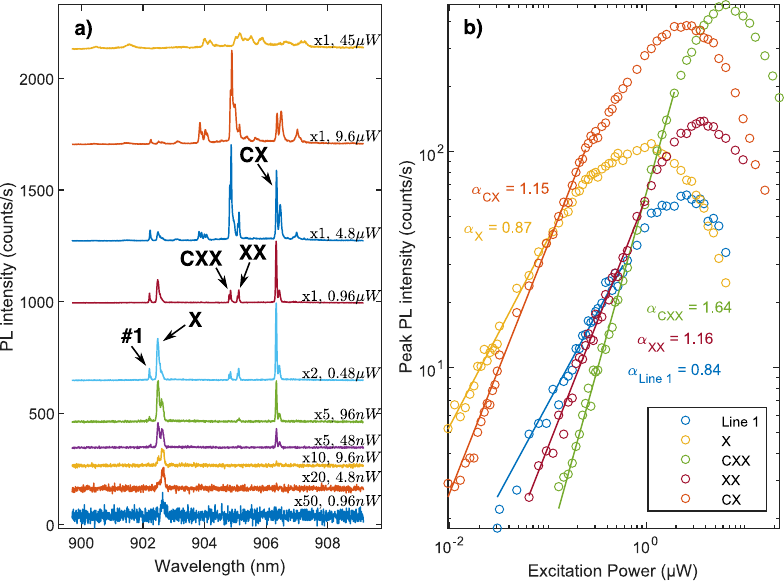}
    \caption{\textbf{Power-dependent photoluminescence of the 5~s InGaAs QD.} \textbf{(a)} High-resolution spectra at different powers measured with a 1800~gr/mm grating. The graphs are staggered to increase visibility. \text{(b)} Peak intensity of each emission line as a function of the excitation power. The intensity $I$ below saturation is fitted with the power law $I \propto Power^{\alpha}$.}
    \MyFigLabel{fig:InGaAs_QD_5s_PPL}
    \end{figure}

    The power evolution of the photoluminescence spectrum gives a first insight into the origin of each emission line (\Cref{fig:InGaAs_QD_5s_PPL}). We find similarities with the 2~s QD case, despite its greater complexity. The neutral exciton (X) and charged exciton (CX) follow both close-to-linear progression again, and saturate when the biexcitons (XX and CXX) become predominant. 

    The power law fit for the charge biexciton returns $\alpha_{CXX} = 1.64$, which is close to a quadratic evolution, as expected for a biexciton state. However, this is not the case for the neutral biexciton line ($\alpha_{XX} = 1.16$); the presence of power-dependent transition rates, such as the charging rate of the QD, in the multi-level dynamics could explain this divergence from the ideal case.

    We also fit a power law to the line \#1, and find a linear dependency, which could indicate that the emission originates from a single-exciton occupancy of the QD \autocite{finleyChargedNeutralExciton2001}.

    \subsection{Polarization-dependent photoluminescence and degree of linear polarization}
    \label[SI]{InGaAs_QD - Polarization-dependent PL of 5s QD}

    \begin{figure}[htbp]
    \includegraphics[width=\textwidth, center]{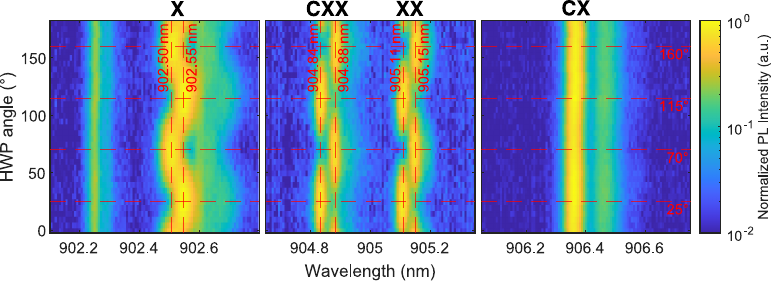}
    \caption{\textbf{Polarization-dependent photoluminescence of the 5~s InGaAs QD.} The exciton (X), biexciton (XX) and charged biexciton (CXX) lines show a fine-structure splitting of $66 \pm 20 ~\mu eV$, while the charged exciton wavelength stays constant with polarization.}
    \MyFigLabel{fig:InGaAs_QD_5s_Pola}
    \end{figure}
    
    The polarization dependency of the exciton line clearly shows a fine-structure splitting (\Cref{fig:InGaAs_QD_5s_Pola}). This result, combined with the power dependency, confirms our choice of labeling for this line. Surprisingly, two other lines exhibit polarization behavior mirroring that of the exciton, which is explained by the presence of a charged biexciton in addition to the neutral biexciton \autocite{poemRadiativeCascadesCharged2010,cadeFineStructureMagnetooptics2006,kettlerNeutralChargedBiexcitonexciton2016}. Such a feature is not atypical, and \Cref{InGaAs_QD - Other examples of 5~s grown InGaAs QD} gives two other examples. We unambiguously differentiate the origins of the two lines using correlation measurements presented later in the text. The charged exciton line does not shift due to its circular polarization as expected due to its circular polarization.

    We use formula \Cref{eq:InGaAs_QD_DLP} to calculate the degree of linear polarization, where $I_H$ ($I_V$) is the photoluminescence intensity when the half-wave plate angle is set to 25\textdegree~(70\textdegree).

    \subsection{Correlation measurements and lifetime measurements}
    \label[SI]{InGaAs_QD - SI:Correlation measurements and lifetime measurements}

    We finally cross-correlate emission from different lines to establish the relation between the transitions (\Cref{fig:InGaAs_QD_5s_Corr}). Despite the complex multi-level dynamic that complicates the interpretation of correlation measurements, we can still make simple observations sufficient to determine the origin of each line without modeling the system using rate equations. 
    
    The cross-correlation of the neutral exciton (X) with the neutral biexciton (XX) \textbf{(a)} shows a strong bunching due to the cascade emission. The bunching is significantly lower in the case of cross-correlation with the charged biexciton (CXX) \textbf{(b)}, due to the additional charging/discharging step. We observed the opposite trend when the biexciton states (XX, CXX) are correlated with the charged exciton (CX) instead \textbf{(c, d)}. The X - CX cross-correlation \textbf{(e)} shows the typical asymmetry coming from the difference between charging and discharging time. Those results unambiguously confirm the origin of the charged exciton, neutral biexciton, and charged biexciton lines. 
    
    To complete the measurements, we cross-correlate the charged biexciton (CXX) with the neutral biexciton (XX) and observe a small bunching \textbf{(f)}. Explaining its origin requires further investigation.
    
    In addition to cross-correlation measurements, we also autocorrelate emission from the exciton states and from the charged biexciton \textbf{(g, i, j)}. The results confirm the single-photon nature of the emission and show the high single-photon purity of the charge exciton emission. 
    
    The antibunching times observed for the excitons (X and CX) are much longer than the ones for the 2~s QD; to check if it is due to a longer lifetime, we measure the time-resolved photoluminescence of the neutral exciton (\Cref{fig:InGaAs_QD_5s_LT}). A strong after-pulse emission is visible with its maximum arriving after 35~ns, indicating that the QD recaptures charges from a long-standing reservoir. We fit the first decay that precedes the after-pulse and measure a characteristic time of 0.85~ns, which is much shorter than the antibunching times and is likely closer to the actual exciton lifetime. The difference between the antibunching times and lifetimes can be explained by the presence of a slow process, such as charging/discharging, within the multi-level dynamics.  

    Additionally, we note that the long timescales observed here are specific to this particular QD, and we measured shorter antibunching times and lifetimes on the nanosecond scale in other 5~s long QDs. \Cref{fig:InGaAs_QD_5s_NW2} presents one example with an antibunching time and lifetime of \mytilde0.65~ns, measured on an exciton state, as the power dependency of its photoluminescence suggests (linearly proportional to the excitation power, see \Cref{fig:InGaAs_QD_5s_NW2_b}).
    
    Finally, we investigated line~1 (\Cref{fig:InGaAs_QD_5s_PPL}), but we could not conclude on the origin of its emission at the present state. We still present our measurements and observations for the sake of completeness. We notice that the cross-correlations of the biexciton states with line 1 \textbf{(k, i)} are very similar to the measurement done with the exciton \textbf{(a, b)}. Moreover, the power dependency of their photoluminescence also looks alike ($ \alpha_{Line \, 1} = 0.84 $ and $ \alpha_{X} = 0.87 $ in \Cref{fig:InGaAs_QD_5s_PPL_b}). This resemblance suggests that line 1 originates from another exciton state.

    \begin{figure}[htbp]
        \centering
        \includegraphics[width=\textwidth]{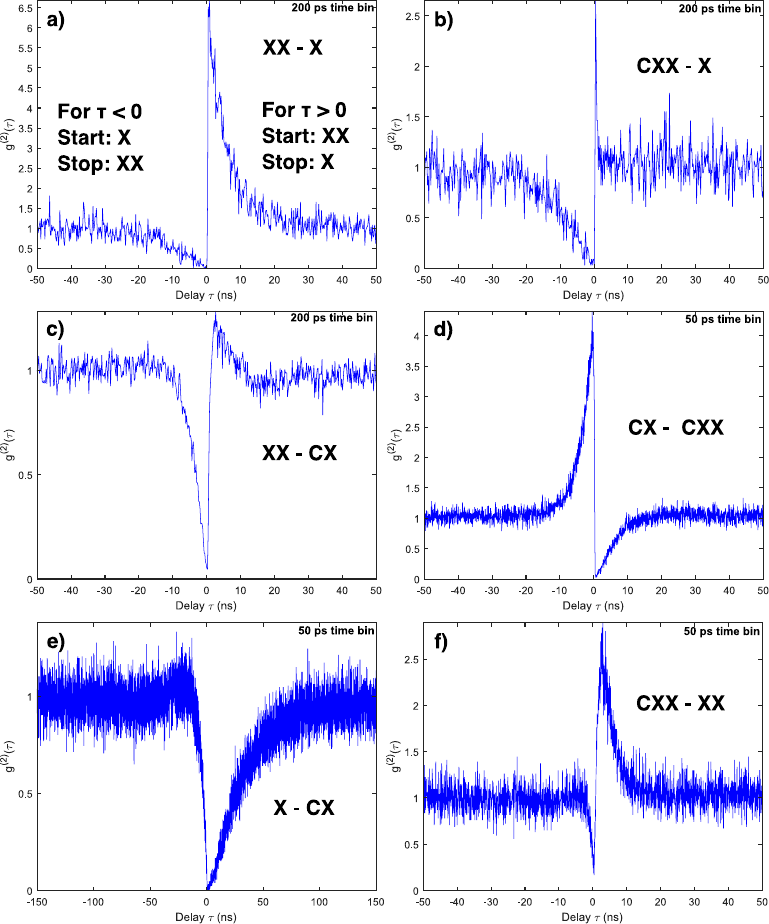}\\
        \medskip
        \emph{(Caption on next page.)}
        \phantomcaption
    \end{figure}

    \begin{figure}[htbp]
        \ContinuedFloat
        \centering
        \includegraphics[width=\textwidth]{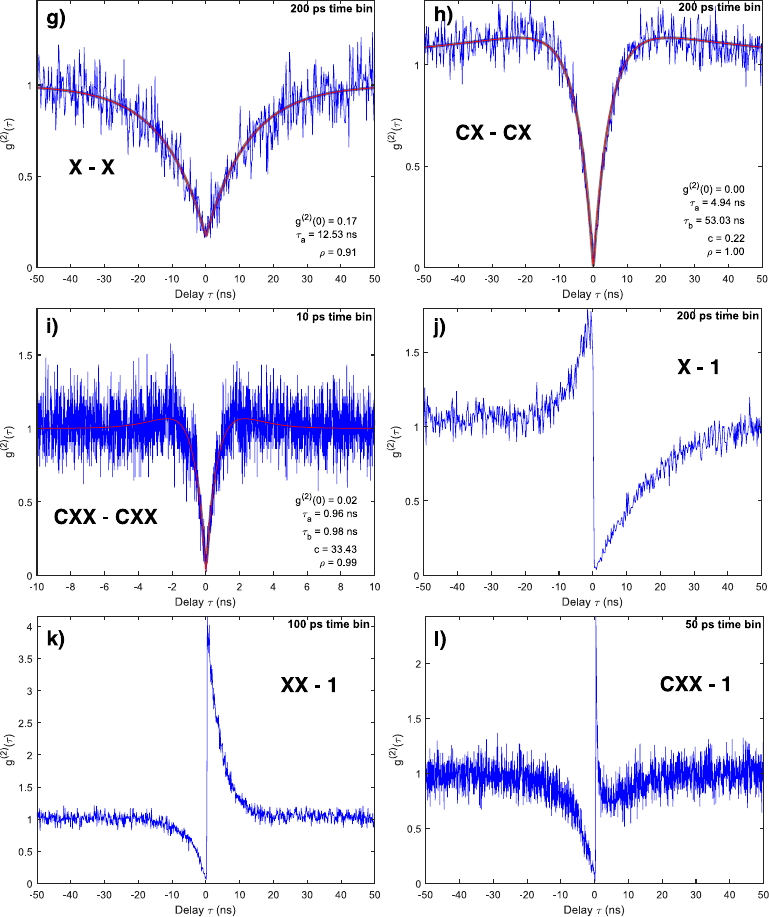}\\
        \medskip
        \emph{(Caption on next page.)}
    \phantomcaption
    \end{figure}

    \begin{figure}[htbp]
        \ContinuedFloat
        \caption{\textbf{Cross-correlation and autocorrelation on the different emission lines of the 5~s QD.} The order in the label of each graph indicates the start and stop channels for positive delay, and the opposite order is valid for negative delay. We set the excitation power of each measurement based on the following condition: \textbf{(e, j)} the exciton (X) countrate is at 80\% of its saturation value, \textbf{(g)} the exciton (X) countrate is at half of its saturation value, \textbf{(h)} the charged exciton (CX) countrate is at half of its saturation value, \textbf{(i)} the charged biexciton (CXX) countrate is at 80\% of its saturation value, \textbf{(a - d, f, k, i)} the two biexciton lines (XX and CXX) have same peak intensity. The time tagger bin size is indicated for each measurement. \textbf{(g - i)} The red line is the result of the fit (transparent gray line) of the g\textsuperscript{(2)} function (formula \Cref{eq:InGaAs_QD_g2_1} or \Cref{eq:InGaAs_QD_g2_2}) deconvoluted by the instrument response function of the measuring setup (Gaussian with a FWHM of 110~ps).}
        \MyFigLabel{fig:InGaAs_QD_5s_Corr}
    \end{figure}

    \begin{figure}[htbp]
    \includegraphics[width=\textwidth, center]{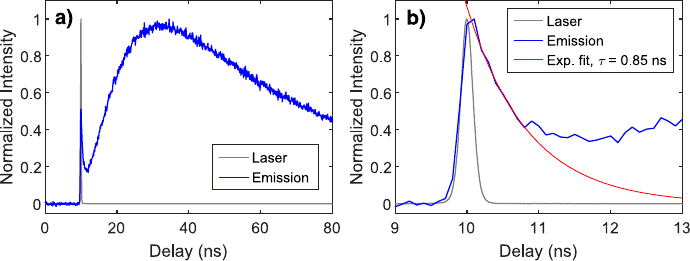}
    \caption{\textbf{Time-resolved photoluminescence of the 5~s QD.} \textbf{(a)} Long-timescale measurement showing a strong recapturing effect. \textbf{(b)} Zoom on the few nanoseconds after the laser pulse. Data points from the first decay are fitted with a decaying exponential with a characteristic time $\tau$.}
    \MyFigLabel{fig:InGaAs_QD_5s_LT}
    \end{figure}

    \begin{figure}[htbp]
    \includegraphics[width=\textwidth, center]{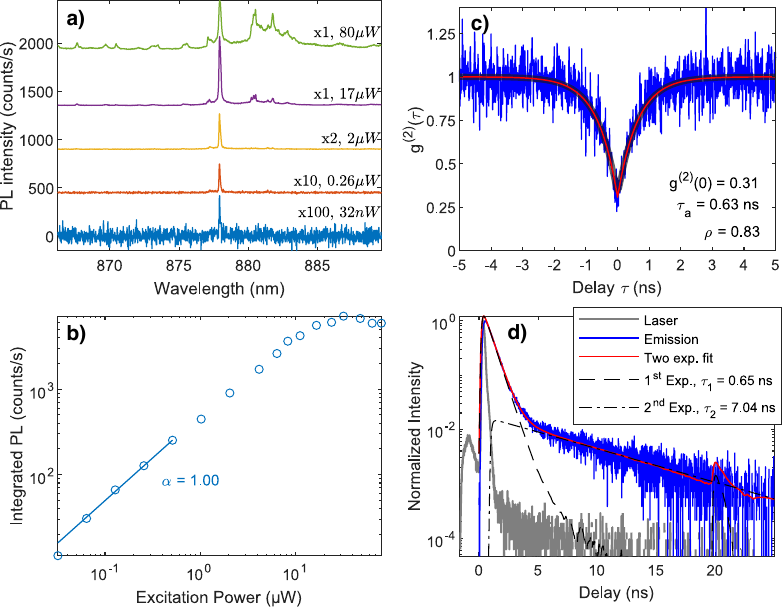}
    \caption{\textbf{Example of 5~s long QD with a nanosecond lifetime.} \textbf{(a)}  High-resolution spectroscopy of the QD emission at different powers and measured with a 1200 gr/mm grating. \textbf{(b)} Integrated number of counts over the FHWM of the emission line at 878~nm. The intensity $I$ below saturation is fitted with the power law $I \propto Power^{\alpha}$. \textbf{(c)} Autocorrelation measurement of the emission line at 878~nm. The red line is the result of the fit (transparent gray line) of the g\textsuperscript{(2)} function (formula \Cref{eq:InGaAs_QD_g2_1}) deconvoluted by the instrument response function of the system (Gaussian with a FWHM of 110~ps). \textbf{(d)} Time-resolved photoluminescence of the emission line at 878~nm. To extract the lifetimes, the data are fitted with decaying exponential functions convoluted with the laser pulse shape (gray line).}
    \MyFigLabel{fig:InGaAs_QD_5s_NW2}
    \end{figure}

\section{Another example of 5~s InGaAs QDs with a charged biexciton state}
\label[SI]{InGaAs_QD - SI:Other examples of 5~s grown InGaAs QD}
\label{InGaAs_QD - Other examples of 5~s grown InGaAs QD}

    \begin{figure}[htbp]
    \includegraphics[width=\textwidth, center]{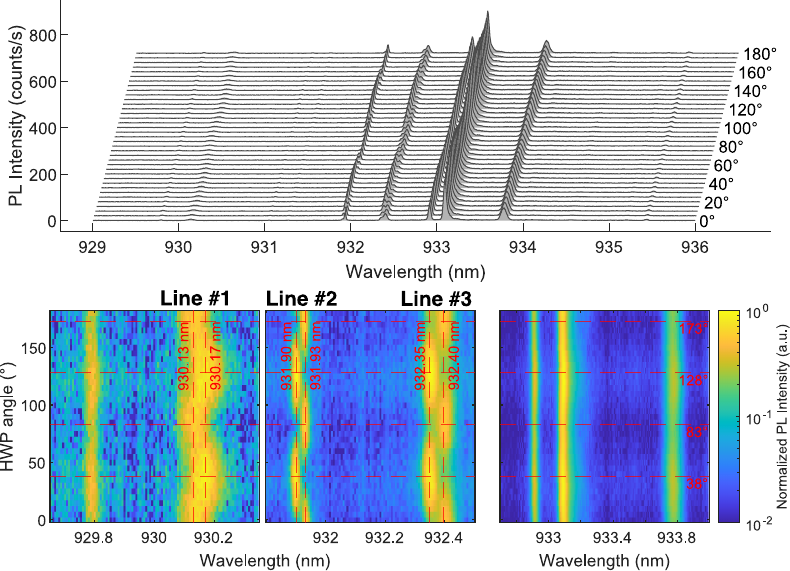}
    \caption{\textbf{Second example of 5~s long QD emission with multiple excitonic lines showing fine-structure splitting.} The top figure presents the full photoluminescence spectrum at different polarization. The graphs are staggered to improve visibility. The bottom figures show the normalized photoluminescence for each region of interest. Line \#1, \#2 and \#3 show a fine-structure splitting of $57 \pm 20 ~\mu eV$.}
    \MyFigLabel{fig:InGaAs_QD_5s_NW6_Pola}
    \end{figure}

    The presence of a charged biexciton state in the emission spectrum of 5~s QDs is not uncommon, and \Cref{fig:InGaAs_QD_5s_NW6_Pola} presents another example. The polarization dependency of line \#1 is mirroring the ones of line \#2 and \#3, which indicates that line \#1 probably originates from a neutral exciton X state, and line \#2 and \#3, from the biexciton states (XX and CXX).

\section{Effect of crystal-phase insertions on the InGaAs QDs}
\label[SI]{InGaAs_QD - SI:Effect of crystal-phase insertions on the InGaAs QDs}

While we could find 2~s InGaAs QDs without any zincblende insertion, all the 5~s InGaAs QDs observed under STEM had at least one crystal-phase insertion. Zincblende and wurtzite phases typically have distinct band structures; therefore, the presence of crystal-phase insertions in the InGaAs QD could impact its electronic states. One of the main questions is then: do crystal-phase insertions split the QD into multiple distinct ones?

To answer, we collect micro-photoluminescence spectra of single nanowires transferred onto a TEM grid and then analyze the crystal structure of their QD under STEM. \Cref{fig:InGaAs_QD_5s_TEM} shows twelve examples of 5~s InGaAs QDs, all presenting at least one crystal-phase insertion. On the photoluminescence spectra, only one group of QD emission lines appears per nanowire. This result strongly suggests that the InGaAs QD is not split into multiple ones when it contains crystal-phase insertions.

Based on this sample of nanowires, the emission from the InGaAs QD varies from 820~nm to ~930~nm. This distribution of wavelength emission is certainly multifactorial, for example, due to fluctuations in size or composition. Here, we identify an additional reason that can play a role — the modulation of the QD band structure by the presence of crystal-phase insertion. This effect has already been observed for embedded QDs in the AlGaAs shell of GaAs nanowires \autocite{francavigliaTuningAdatomMobility2018}.

\section{Absence of emission originating from the thin InGaAs layer}
\label[SI]{InGaAs - SI:Absence of emission from the thin InGaAs layer}

We can also conclude on the effect of the thin InGaAs layer from \Cref{fig:InGaAs_QD_5s_TEM}. If the InGaAs layer emits light, its emission should be blueshifted compared to the InGaAs QD emission, because its most confined dimension is shorter than the QD size (stronger quantum confinement). However, we do not observe emission in the spectral range between the AlGaAs core-shell and the InGaAs QD, which indicates that the InGaAs layer is not emitting. The quenching of its emission can be explained by its thickness, which is too thin to localize charge carriers\autocite{lambkinThermalQuenchingPhotoluminescence1990}.

\begin{figure}[htbp]
    \centering
    \includegraphics[width=0.8\textwidth]{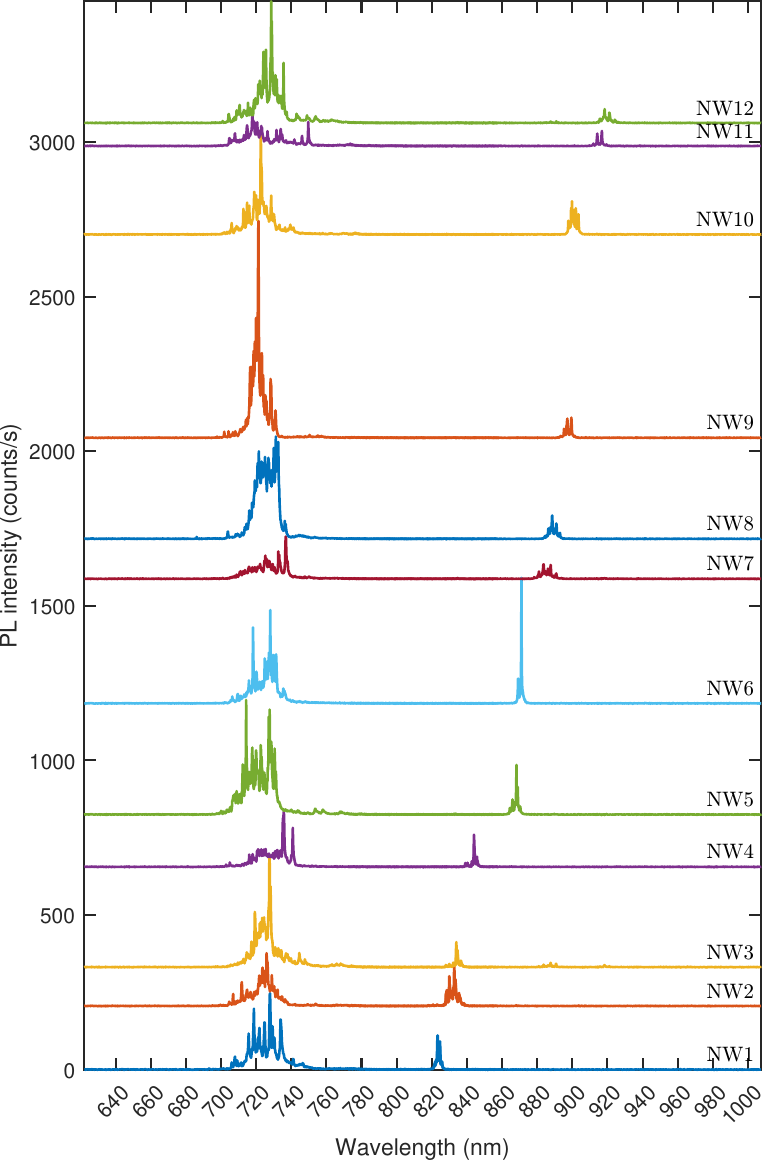}\\
    \medskip
    \emph{(Caption on next page.)}
    \phantomcaption
\end{figure}

\begin{figure}[htbp]
    \centering
    \ContinuedFloat
    \includegraphics[width=0.8\textwidth]{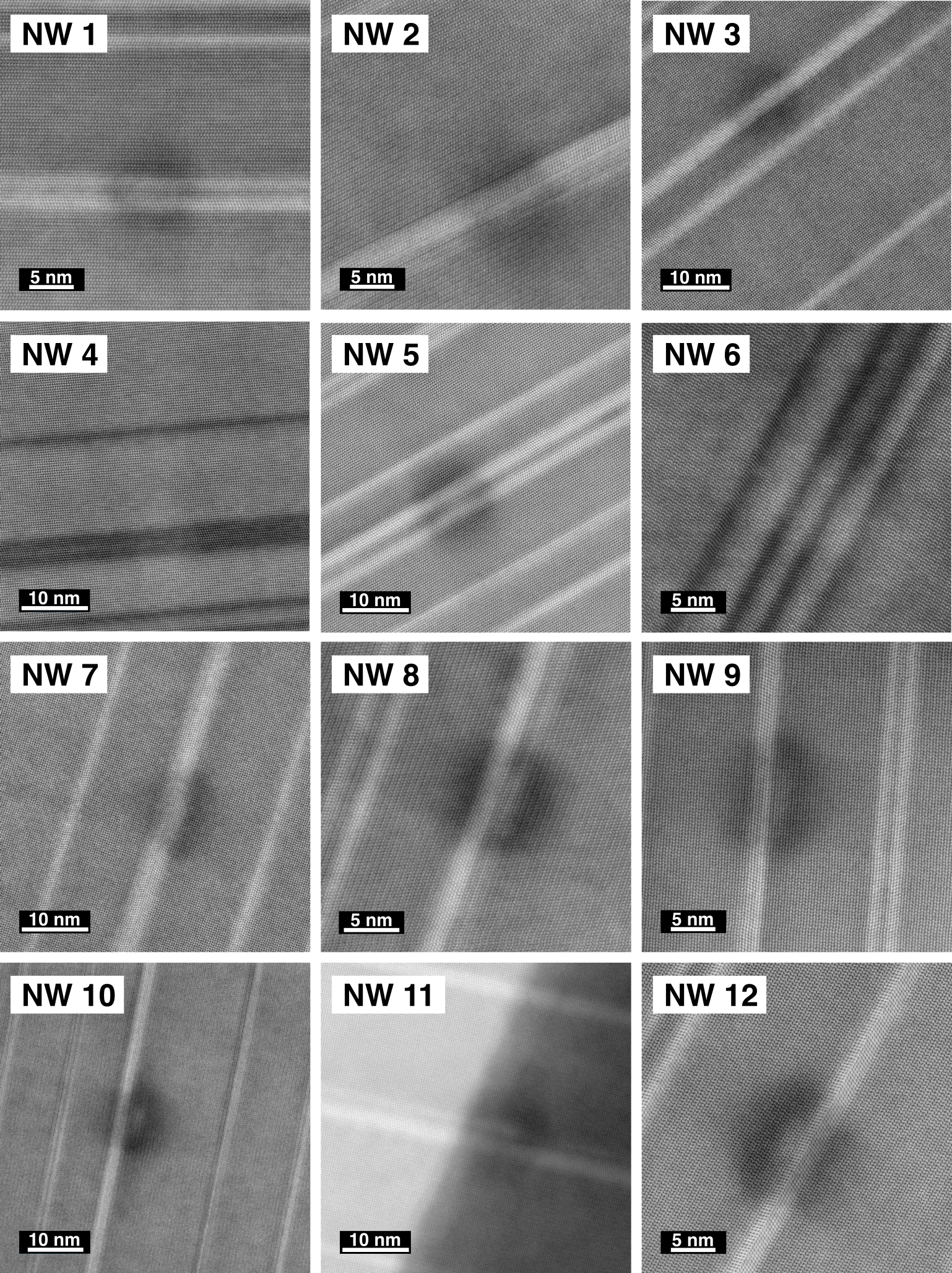}
    \caption{\textbf{Photoluminescence spectra and corresponding STEM images of 5~s InGaAs QDs.} All the photoluminescence spectra were collected at the same excitation power ($3.2~\mu W$). AlGaAs core-shell emits from 700 to 750~nm, and the InGaAs QDs emit above 820~nm. All images have been acquired by bright-field STEM (BF-STEM) except for NW 4 and 6 which are made by HAADF-STEM.}
    \MyFigLabel{fig:InGaAs_QD_5s_TEM}
\end{figure}

\section{Structural and compositional analysis of the InAs segments grown on top of the AlGaAs nanowires}
\label[SI]{InGaAs - SI:Compositional analysis of the InAs segments grown on top of the AlGaAs nanowires}

The compositional purity of the InAs segments grown on top of AlGaAs nanowires presented in \Cref{fig:InGaAs_QD_Fig4_cd} of the main text was confirmed by EDX measurement. \Cref{fig:InGaAs_QD_EDX_InAs_a} presents the measured spectrum, which shows the presence of indium and arsenic, and the absence of aluminum and gallium. Moreover, the InAs segment has a pure wurtzite crystal structure as shown in \Cref{fig:InGaAs_QD_EDX_InAs_b}, which is an enlargement of \Cref{fig:InGaAs_QD_Fig4_d} in the main text.

\begin{figure}[htbp]
    \includegraphics[width=\textwidth, center]{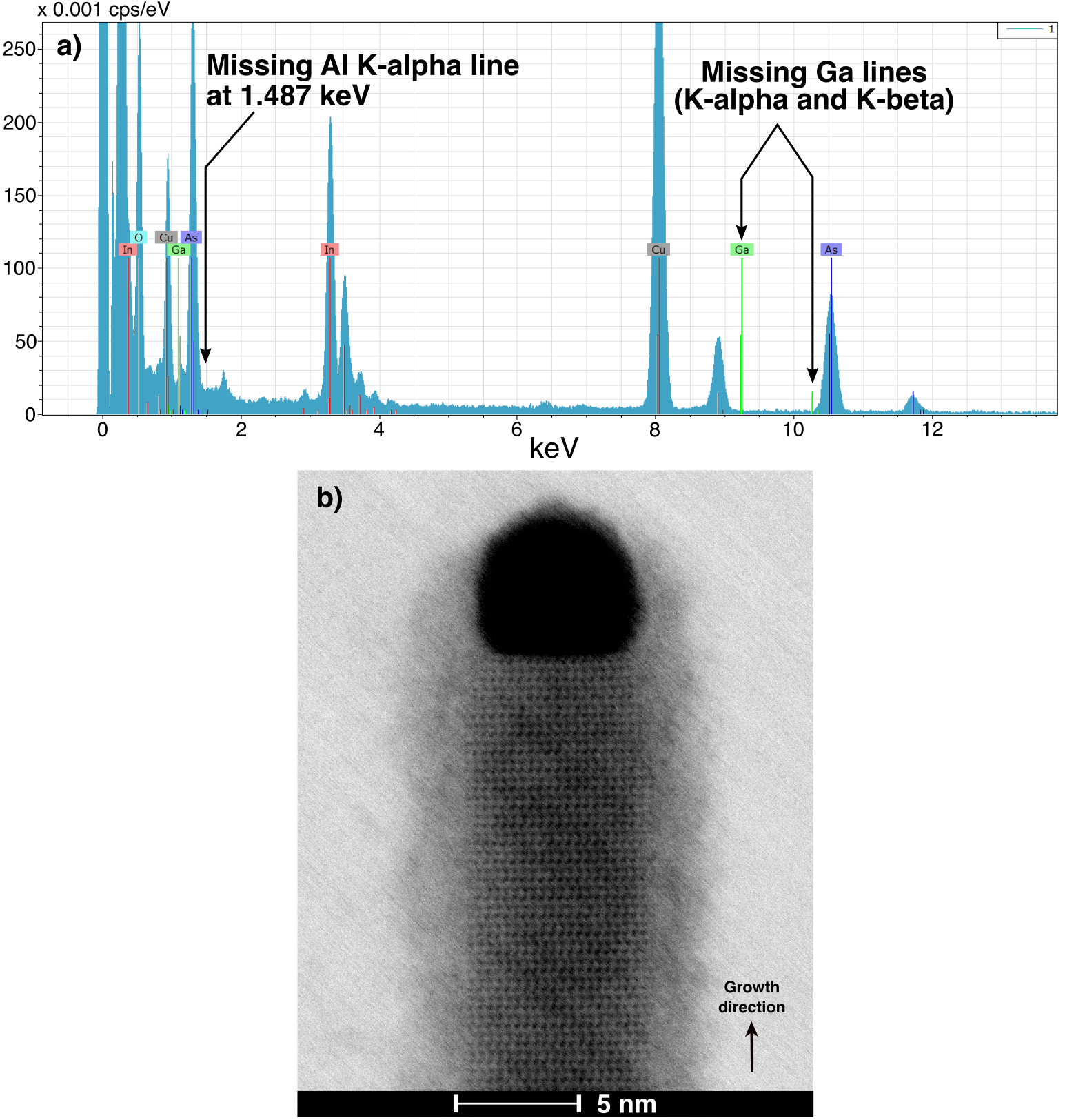}
    \caption{\textbf{Structural and compositional analysis of the InAs segments grown on top of the AlGaAs nanowires.} \textbf{(a)} EDX spectrum measured on an InAs segment grown on top of an AlGaAs nanowire, showing its compositional purity with the absence of Al and Ga emission (Al K-alpha, Ga K-alpha, and Ga K-beta emission lines). \textbf{(b)} High-resolution BF-STEM showing the pure wurtzite structure of the InAs segment.}
    \MyFigLabel{fig:InGaAs_QD_EDX_InAs}
\end{figure}

\clearpage 
\printbibliography

\end{document}